\documentclass{iopart}

\usepackage{graphicx}
\usepackage{dcolumn}
\usepackage{bm}
\usepackage{iopams}
\usepackage{color}

\newtheorem{remark}{{\bf Remark}}

\begin{document}
\title[Beltrami disk-jet flow]{Generalized Beltrami flow ---
a model of thin-disk and narrow-jet system}

\author{Z. Yoshida$^1$ and N. L. Shatashvili$^{2,3}$}%

\address{$^1$Graduate School of Frontier Sciences, The University of Tokyo,
Chiba 277-8561, Japan
\\
$^2$Faculty of Exact and Natural Sciences, Javakhishvili Tbilisi State University, Tbilisi 0128, Georgia
\\
$^{3}$Andronikashvili Institute of Physics, Javakhishvili Tbilisi
State University, Tbilisi 0177, Georgia
}

\begin{abstract}
In the vicinity of a massive object of various scales (ranging from young stars
to galactic nuclei), mass flow creates a spectacular structure combining
a thin disk and collimated jet.
Despite a wide range of scaling parameters (such as Reynolds number, Lundquist number,
ionization fractions, Lorentz factor, etc.),
they exhibit a remarkable similarity that must be dictated by a universal principle.
A generalized Beltrami condition has been formulated as a succinct representation
of such a principle.
The singularity at the center of the Keplerian rotation 
forces the flow to align with the
``generalized vorticity'' (including the effect of localized density and
finite dissipation) which appears as an axle penetrating the disk,
i.e. the jet is a Beltrami flow.
Based on the Beltrami flow model, an analytical expression of
a disk-jet system has been constructed by the method of similarity solution.
\end{abstract}

\maketitle

\section{Introduction}

\begin{figure}
\begin{center}
\includegraphics[scale=0.7]{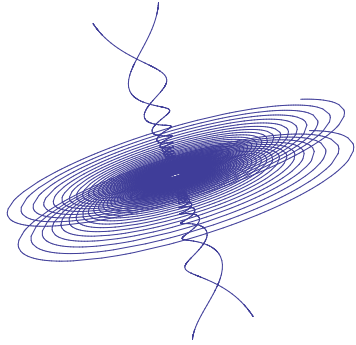}
\caption{
In a disk-jet system,
the accreting flow and jet align parallel to a \emph{generalized vorticity}.
This figure shows the streamlines of a ``generalized Beltrami flow''
to be constructed in Sec.\,\ref{sec:similarity-solution}
}
\label{fig:streamline}
\end{center}
\end{figure}

The combination of a thin disk and collimated jet is a common
structure that is created in the vicinity of a massive object
\cite{bib:bland1-2,bib:begelman4,bib:livio,bib:ferrari,bib:livio2}.
Beneath a large variety of scales, constituents, and local
processes of such systems, there must be a simple and universal
principle that dictates the remarkably similar geometry;
see Fig.\,\ref{fig:streamline}.
Here we show that the collimated structure of jet is a natural
consequence of the \emph{alignment} of the flow velocity and the
vorticity, i.e. so-called \emph{Beltrami condition} determines the structure.
On a Keplerian thin disk, the
vorticity becomes a vertical vector with a magnitude $\propto
r^{-3/2}$ ($r$ is the radius from the center of the disk), which
appears as a spindle of the disk. Then, the alignment is the
unique solution for avoiding singularity of Coriolis force near the center.
However, we need to generalize the
``vorticity'' to deal with the
strong heterogeneity of the disk-jet system,
as well as to account for the dissipation that causes accretion.
The mission of this study is to
formulate an appropriate \emph{generalized vorticity}
to which the disk-jet flow aligns.

Let us start by a short review of the Beltrami vector field
which has wide applications as a model
of various \emph{vortex structures} found in nature.
The Beltrami condition, demanding the alignment of flow and its
vorticity, forces the total \emph{energy density} (consisting of thermal energy, kinetic energy,
and other energies of coupled fields such as gravitational or electromagnetic) to
distribute homogeneously (so called Bernoulli condition); the
Beltrami-Bernoulli condition, thus, fits the notion of
``relaxed state.''
While a Beltrami field was discovered
by many researchers as a particular type of equilibrium
state\,\cite{bib:ChandrasekharKendall1957,bib:low}, or \emph{free decay}
solution\,\cite{bib:vanKampen}, in fluids or
plasmas, its relation to the \emph{helicity}
was noticed in the study of ``force-free magnetic
fields'' in plasmas; Woltjer\,\cite{bib:Woltjer1958} invoked the
helicity as a constraint in minimizing the magnetic energy;
Taylor\,\cite{bib:JBT1974} considered that the ``relaxed
state'' is the energy minimizer under the lugged constancy of the total helicity;
the corresponding Euler-Lagrange equation
becomes the eigenvalue problem of the curl operator;
see \cite{bib:yoshida-giga} for the mathematical characterization of the curl operator
and its eigenfunctions.
We can actually observe Taylor relaxed states on various experiments\,\cite{bib:JBT-RMP},
as well as in some astronomical systems;
e.g. \cite{bib:Heyvaerts-Priest1984,bib:Kusano1995}.
The helicity constraint causes a finite \emph{vorticity}
in the relaxed state, resulting in interesting topological properties
of field-lines\,\cite{bib:Moffatt}.
In the context of Hamiltonian mechanics,
the helicity is regarded as a Casimir element representing the
defect of the governing symplectic geometry\,\cite{bib:Morrison};
a Beltrami field is an equilibrium point on a \emph{helicity leaf}.

The Beltrami fields constitute an interesting, widely-applied class of
vectors (or axial vectors) with ``twisted'' field-lines;
e.g. \cite{bib:Cantarella,bib:Dritschel,bib:Tang}.
Not only for describing equilibrium states,
they are applied to analyze waves\,\cite{bib:Yoshida1991,bib:Gonzalez},
instabilities\,\cite{bib:Yoshida2003}, and turbulence\,\cite{bib:Montgomery,bib:Ito,bib:Yang} (here we can
cite only a short list of references).

A variety of generalizations have been proposed.
Including the cross helicity as an
additional constraint, we obtain a flow parallel to
the magnetic field and the resultant hydrodynamic pressure balancing with a
static pressure\,\cite{bib:Sudan1979}.
The two-fluid (Hall MHD) formulation elucidates
a fundamental structure in the coupling of flow and magnetic field in terms of the
\emph{canonical vorticities}\,\cite{bib:MahajanYoshida1998};
the simultaneous ion and electron Beltrami conditions
yields the \emph{double Beltrami fields},
which have various applications in both laboratory and
astrophysical plasmas; e.g. \cite{bib:Ohsaki,bib:Mahajan-Nana,bib:Yoshida2010}.
Another generalization is made by including the coupling of vortex and compressible motions;
by boosting such generalized Beltrami fields, we obtain
modulating nonlinear Alfv\'en waves\,\cite{bib:Yoshida2011}.

This work extends the scope of Beltrami fields to show that
the disk-jet system is a ``generalized'' Beltrami vortex\,\cite{bib:Shatashvili};
the generalization is made by introducing a new \emph{generalized vorticity}
that combines the vorticity of a ``reduced momentum'' ---the reduction is to account for
a viscous dissipation as well as to subtract the centrifugal
force of the Keplerian rotation.

\section{Generalized vorticity and Beltrami condition}
\label{sec:formulation}

To demonstrate how the alignment condition arises and how it
determines the structure of a thin disk and
narrowly-collimated jet, we invoke a simple
model of neutral fluid; for a generalization to magneto-fluid, see
Appendix A.
Let $ \bm{P} = \rho \bm{V}$ denote the mechanical momentum density, where
$\rho$ is the mass density and $\bm{V}$ is the (ion) flow
velocity. The momentum equation read as
\begin{equation}
\partial_t \bm{P} + \nabla\cdot(\bm{V}\bm{P})
= -\rho\nabla\phi - \nabla p -\nabla\cdot\bm{\Pi} ,
\label{momentum-i-1}
\end{equation}
where $\phi$ is the gravity potential, $p$ is the scalar pressures, and $\bm{\Pi}$
is the (effective) viscosity tensor.
The variables are normalized as follows: We choose a
representative flow velocity $V_0$ and a mass density $\rho_0$ in
the disk, and normalize $\bm{V}$ and $\rho$ by these units. The
energy densities $\rho\phi$
(gravitational) and $p$ (thermal) are normalized by the unit
kinetic energy density ${\cal{E}}_0:=\rho_0 V_0^2/2$.
The independent variables (coordinate $\bm{x}$ and time $t$) are
normalized by the system size $L_0$ and the corresponding transit
time $T_0=L_0/V_0$.

We consider stationary solutions; putting
$\partial_t=0$ in (\ref{momentum-i-1}), we obtain
\begin{equation}
\nabla\cdot(\bm{V}\bm{P}) =
-\rho\nabla\phi - \nabla p - \nabla\cdot\bm{\Pi}.
\label{momentum-i-1-scaled}
\end{equation}
In order to derive a term that balances with the viscosity term, we
decompose the ``inertia term'' [the left-hand side of
(\ref{momentum-i-1-scaled})] as follows: we first write
\begin{equation}
\rho = \rho_1 \rho_2 , \label{decomposition1}
\end{equation}
and denote
\begin{equation}
\bm{P}_1 := \rho_1\bm{V}, \quad \bm{P}_2 := \rho_2 \bm{V}.
\label{decomposition2}
\end{equation}
Using these variables, we may write
\begin{equation}
\nabla\cdot(\bm{V}\bm{P})
=\nabla\cdot(\rho\bm{V}\bm{V})
=(\nabla\cdot\bm{P}_1)\bm{P}_2 + (\bm{P}_1\cdot\nabla)\bm{P}_2.
\label{decomposition3}
\end{equation}
In the conventional formulation of fluid mechanics, we choose
$\rho_2=1$ and $\rho_1=\rho$.  Then,
using the mass conservation law $\partial_t \rho +
\nabla\cdot\bm{P}=0$, we can rewrite the left-hand side of
(\ref{momentum-i-1}) as $\rho[\partial_t\bm{V}+(\bm{V}\cdot\nabla)\bm{V}]$.
In the preset analysis, however, we choose a different separation of the
inertia term in order to match $(\nabla\cdot\bm{P}_1)\bm{P}_2$
with the viscosity term.
By a reduced $\rho_2$ ($<1$), we will define a \emph{generalized vorticity}
of a \emph{reduced momentum}.
Multiplying $(\rho_2/\rho_1)$ on both sides of (\ref{momentum-i-1-scaled}), we obtain
(assuming a barotropic relation, we put $\nabla p=\rho\nabla h$ with an enthalpy $h$)
\begin{equation}
\bm{P}_2\times\bm{\Omega}_2 = \frac{1}{2}\nabla P_2^2 + \rho_2^2
\nabla\,(\phi + h)
+ \frac{\rho_2}{\rho_1}\,[(\nabla\cdot\bm{P}_1)\bm{P}_2 + \nabla\cdot\bm{\Pi}] ,
\label{momentum-2}
\end{equation}
where
\begin{equation}
\bm{\Omega}_2
:= \nabla\times\bm{P}_2
\label{g-vorticity}
\end{equation}
is a \emph{generalized vorticity}.

In the next section, we will show that a generalized Beltrami condition,
demanding that $\bm{\Omega}_2$ parallels $\bm{P}_2$
(thus, (\ref{momentum-2}) holds with both sides being zero),
is a unique recourse to avoid singular energy densities in a disk-jet geometry.

\section{Beltrami Model of Disk-Jet System}
\label{Sec:Disk-Jet}

Now we consider an axisymmetric ($\partial_\theta =0$ in the
$r$-$\theta$-$z$ coordinates) disk-jet system.
A massive central object produces $\phi = -MG/r$
(we neglect the mass in the disk and jet). In the disk,
$\bm{V} \approx V_\theta \bm{e}_\theta$ with
the Keplerian velocity $V_\theta \propto r^{-1/2}$.
Then, $\nabla\times\bm{V} \propto r^{-3/2} \bm{e}_z$.
The momentum is strongly
localized in the thin disk, and the vorticity diverges near the axis.
This particular configuration
poses strong constraints on the force balance equation (\ref{momentum-2}),
allowing only a special class of solutions to exist;
following conclusions are readily deducible.

\subsection{Balance of viscosity force and partial inertia}
\label{subsec:partial_inertia}
In the disk, a radial flow (much smaller than
$V_\theta$) is caused by the viscosity
(a finite dissipation breaks the conservation of the angular momentum and enables
the flow to cause accretion).

Since the flow $\bm{V}$ is primarily in the azimuthal ($\theta$) direction,
the viscosity force can be approximated as
(under assumption of the azimuthal symmetry, $\nabla\cdot\bm{V}\approx0$)
\begin{equation}
-\nabla\cdot\bm{\Pi} \approx -\nabla\times (\rho\eta\nabla\times\bm{V}),
\label{viscosity_term}
\end{equation}
where $\eta$ is the shear viscosity coefficient.
In a Keplerian thin disk, where $\bm{V}\approx V_0 r^{-1/2}\bm{e}_\theta$, and
$\nabla(\rho\eta)$ is approximately vertical,
we may estimate
\begin{equation}
-\nabla\cdot\bm{\Pi} \approx
-\rho \eta \nabla\times(\nabla\times\bm{V}) =
- \rho \eta V_0 \frac{3}{4} r^{-5/2} \bm{e}_\theta.
\label{viscosity_term-2}
\end{equation}
Hence, we may write $-\nabla\cdot\bm{\Pi}= -\nu \bm{P}$ with a positive coefficient $\nu(r)$,
i.e., the viscosity force is primarily in the azimuthal (toroidal)
direction, which can be balanced by the term
$(\rho_2/\rho_1)\,(\nabla\cdot\bm{P}_1)\bm{P}_2$ that has been
extracted from the inertia term;
see (\ref{decomposition3}). Using the steady-state mass
conservation law $\nabla\cdot\bm{P}=0$, we observe
\[
\rho_1^{-1} \nabla\cdot\bm{P}_1
=  \rho_2 \bm{V}\cdot\nabla\rho_2^{-1} = -
\bm{V}\cdot\nabla\log\rho_2 .
\]
Hence, the  balance of the viscosity and the partial inertia term
demands
\begin{equation}
\bm{V}\cdot\nabla\log\rho_2 = \nu,
\label{friction-balance}
\end{equation}
which determines the parameter $\rho_2$.
(Since $\partial_\theta=0$, we can integrate (\ref{friction-balance}) for
$\rho_2$ along the streamline of $\bm{V}$ on the poloidal $r$-$z$ plane.)

The remaining part $\rho_1$ of the density is determined by the mass
conservation law:
By $\nabla\cdot\bm{P} =\nabla\cdot(\rho\bm{V})
= \rho_2\bm{V}\cdot\nabla\rho_1 + \rho_1\bm{V}\cdot\nabla\rho_2
+ \rho_1\rho_2\nabla\cdot\bm{V}
= 0$
and (\ref{friction-balance}), we obtain a relation
\begin{equation}
\bm{V}\cdot\nabla\log\rho_1 = -\nabla\cdot\bm{V}-\nu.
\label{friction-balance'}
\end{equation}

\subsection{Beltrami condition}
\label{subsec:Beltrami}
Near the axis, the poloidal component of the flow
begins to have an appreciable vertical ($z$) component
---this is the place where the jet is created;
we are going to unearth the mechanism that collimates the flow.

After balancing the third and fourth terms in
(\ref{momentum-2}), the remaining terms do not have a
toroidal (azimuthal) component.
In fact, the right-hand-side
gradient terms have only poloidal components, and
hence, the left-hand-side $\bm{P}_2\times\bm{\Omega}_2$ must not have a toroidal component
(to put it in another way, we have extracted the partial inertia term
$-(\rho_2/\rho_1)(\nabla\cdot\bm{P}_1)\bm{P}_2$ from the total
inertia to separate the toroidal component).
The vorticity $\bm{\Omega}_2$ includes a singular factor
$\nabla\times\bm{V} \propto r^{-3/2}\bm{e}_z$.
To eliminate the divergence of $\bm{P}_2\times\bm{\Omega}_2$ near the axis,
$\bm{P}_2$ must \emph{align} to $\bm{\Omega}_2$,
i.e., the \emph{Beltrami condition}
\begin{equation}
\bm{\Omega}_2 = \lambda \bm{P}_2
\label{Beltrami1}
\end{equation}
must be satisfied, where $\lambda$ is a certain scalar function.
The flow $\bm{V}=\bm{P}_2/\rho_2$ is, therefore, collimated by the
\emph{generalized vorticity} $\bm{\Omega}_2$ creating a jet.

\begin{remark}
\label{remark:Keplerian-velocity}
Here the essential part of the Beltrami condition is its poloidal component,
which dictates the poloidal flow so as to eliminate the toroidal component of the inertia term $\bm{P}_2\times\bm{\Omega}_2$.
As for the toroidal flow $V_\theta\bm{e}_\theta$,
which yields primarily a radial (centrifugal) inertia force, the ``Beltrami condition''
brings about an extra constraint.
In fact, if we were to use the conventional vorticity
$\bm{\Omega}=\nabla\times\bm{V}$ (i.e., if $\rho_2=1$) and estimate the
centrifugal force of the Keplerian flow $\bm{V}=V_0r^{-1/2}\bm{e}_\theta$,
the term $\bm{P}\times\bm{\Omega}$ contributes a half of
the total centrifugal force $(\bm{V}\cdot\nabla)\bm{P} = \rho V_0^2/r$,
while the term $\nabla V^2/2$ on the right-hand side of (\ref{momentum-2})
contributes the remaining half; combining these two terms, we obtain
the right balance with the gravity $-\rho MG/r^2$.
Hence, the conventional Beltrami condition $\bm{P}\times\bm{\Omega}=0$
would lead to an inadequate estimate of the toroidal flow.
However, our \emph{generalized Beltrami condition},
based on the generalized vorticity $\bm{\Omega}_2$ (including $\rho_2\neq1$),
can be made consistent with the Keplerian velocity;
see Sec.\,\ref{sec:similarity-solution}.
\end{remark}

\subsection{Bernoulli condition}
\label{subsec:Bernoulli}
When the Beltrami condition eliminates the left-hand side of
(\ref{momentum-2}), the remaining potential forces must
balance to achieve the \emph{Bernoulli
condition}\,\cite{bib:MahajanYoshida1998} that reads as
\begin{equation}
\frac{1}{2\rho_2^2}\nabla P_2^2 + \nabla( \phi + h)
=\nabla \left( \frac{1}{2}V^2 + \phi + h \right) + V^2 \nabla
\log\rho_2
= 0. \label{Bernoulli-1}
\end{equation}

The system of determining equations is summarized as follows: By
(\ref{friction-balance}), we determine $\rho_2$
for a given $\nu$. This equation involves $\bm{V}=\bm{P}/\rho$
that is governed by the Beltrami equation (\ref{Beltrami1}).
After determining $\bm{V}$ and $\rho_2$, we can solve the
Bernoulli equation (\ref{Bernoulli-1}) to determine $h$
(the gravitational potential is approximated by $\phi=-MG/r$).

\section{Parameterization by Clebsch potential}
\label{Sec:2D-formulation}

We may rewrite the determining equations
(\ref{friction-balance})-(\ref{Bernoulli-1})
in a succinct form by invoking the Clebsch parameterization.
In an axisymmetric geometry, the divergence-free vector $\bm{P}$
may be parameterized as
\begin{equation}
\bm{P} = \nabla\psi\times\nabla\theta + I \nabla \theta,
\label{Grad-form-1}
\end{equation}
where $I = \rho r V_\theta$.  Both $\psi$ and $I$ do not depend on
$\theta$. Since $\bm{P}\cdot\nabla\psi=0$, the level sets
(contours) of $\psi$ are the streamlines of $\bm{P}$ (or those of
$\bm{V}=\bm{P}/\rho$).
In a disk region, $rV_\theta \propto r^{1/2}$,
while $\rho$ is a strongly localized function with respect to $z$.

Substituting (\ref{Grad-form-1}) into
(\ref{friction-balance}) yields
\begin{equation}
\nu = \frac{1}{\rho} \bm{P}\cdot\nabla \log \rho_2
= \frac{1}{r\rho}\{\log\rho_2,\psi\},
\label{friction-balance-2}
\end{equation}
where $\{a,b\} := (\partial_r b) (\partial_z a) - (\partial_r
a)(\partial_z b)$. For a given set of $\bm{P}$, $\rho$ and $\nu$,
we can solve (\ref{friction-balance-2}) to determine
$\rho_2$, as well as $\rho_1=\rho/\rho_2$ that is consistent to
(\ref{friction-balance'}).

Substituting (\ref{Grad-form-1}) into the Beltrami condition (\ref{Beltrami1}),
we obtain,
from the toroidal component,
\begin{equation}
\lambda\rho_1^{-1} \nabla\psi= \nabla \left(\rho_1^{-1}I \right) ,
\label{Beltrami-I-unmag}
\end{equation}
implying that $\rho_1^{-1}I =: I_2 = I_2(\psi)$ and
$\lambda\rho_1^{-1} = I_2'(\psi) $ (we denote
$f'(\psi)=\rmd f(\psi)/\rmd\psi$),
and, from the poloidal component,
\begin{equation}
{\cal L} \psi - \nabla\psi\cdot\nabla\log\rho_1 = - \rho_1^2 I_2'(\psi) I_2(\psi) ,
\label{Beltrami3'}
\end{equation}
where ${\cal L}\psi := r \partial_r(r^{-1}\partial_r \psi) +
\partial_z^2 \psi$.  This elliptic partial differential equation determines
the poloidal-momentum Clebsch potential $\psi$.

The Beltrami condition has decoupled the gradient forces from the momentum
equation, which must balance separately ---the Bernoulli
condition (\ref{Bernoulli-1}) which now reads as
the determining equation of the enthalpy:
\begin{equation}
\nabla h =
-\nabla \left[ \frac{1}{2r^2\rho^2}\left(|\psi|^2+ I^2 \right) + \phi \right]
-  \frac{1}{r^2\rho^2}\left(|\psi|^2+ I^2 \right) \nabla \log\rho_2.
\label{Bernoulli-2}
\end{equation}


\section{Analytic Similarity Solution}
\label{sec:similarity-solution}

\subsection{A similarity solution modeling disk-jet structure}

In this section, we construct a \emph{similarity solution} of the
model (\ref{Beltrami3'}), which describes a
fundamental disk-jet structure.
We define
\begin{equation}
\tau := \frac{z}{r}
\quad (r>0),
\label{similarity-0}
\end{equation}
and an orthogonal variable ($\nabla\tau\cdot\nabla\sigma=0$)
\begin{equation}
\sigma := \sqrt{r^2 + z^2} \ . \label{similarity-0'}
\end{equation}
In the thin disk region, we may approximate $\sigma\sim r$, while in the
narrow jet region, $\sigma\sim z$.
The system is mirror symmetry with respect to the $z=0$ plane, and
the axes $r=0$ and $z=0$ are left as singularities.
We consider $\psi$ such that
\begin{equation}
\psi = \psi(\tau) = -J \tau^p  -D \tau^{-q},
\label{similarity-1}
\end{equation}
where $J$ and $p$ ($D$ and $q$) are positive constants, which
control the strength of the jet (disk) flow.

\begin{figure}
\begin{center}
\includegraphics[scale=0.7]{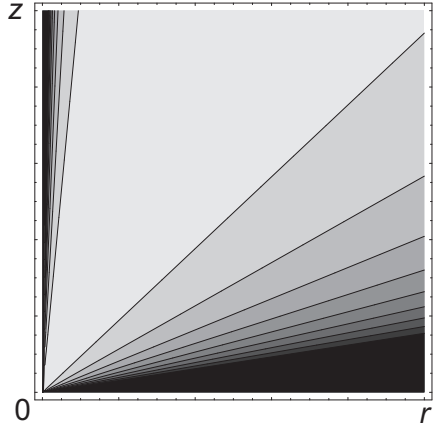}
\caption{ The momentum field (contours of $\psi$ that describe the
streamlines of the poloidal component of $\bm{P}$) of the
similarity solution (with $D=1$, $p=1$, $J=0.1$ and $q=1$). }
\label{fig:flow}
\end{center}
\end{figure}

As shown in Fig.\,\ref{fig:flow}, this $\psi$ models a disk-jet flow.
The level sets of $\psi$ (hence, those of $\tau$) are the
streamlines of $\bm{P}$. On the other hand, $\sigma$ serves as the
coordinate directed parallel to the streamlines. We assume that
$\rho_1$ is written as
\begin{equation}
\rho_1(\tau,\sigma) = \rho_\perp(\tau) \rho_\parallel(\sigma),
\label{similarity-1'}
\end{equation}
and, then, $\log\rho_1= \log\rho_\perp(\tau)+\log\rho_\parallel(\sigma)$.

Let us see how the stream function $\psi$ defined by (\ref{similarity-1})
satisfies the determining equations (\ref{friction-balance-2}), (\ref{Beltrami3'}), and
(\ref{Bernoulli-2}), i.e., we determine all other
fields $I_2(\psi)$, $\rho_1$, $\rho_2$, $\nu$ and $h$ that allow
this $\psi$ to be the solution. For arbitrary $f(\tau)$ and
$g(\tau)$, we observe
\[
{\cal L} f =
\frac{1}{r^2} \left[ (\tau^2+1) f'' + 3\tau f' \right] ,
\]
\[
\nabla f \cdot \nabla g
= \frac{1}{r^2} (\tau^2+1)f' g' .
\]
Hence, the left-had side of (\ref{Beltrami3'}) is (denoting
$g(\tau):=\log\rho_\perp(\tau)$)
\begin{equation}
{\cal L} \psi - \nabla \psi \cdot \nabla \log \rho_1
= \frac{1}{r^2} \left[ (\tau^2+1) \psi'' + 3\tau \psi' - (\tau^2+1)g'\psi' \right].
\label{similarity-2}
\end{equation}
For this quantity to balance with the right-hand side of
(\ref{Beltrami3'}),
$r^{-2}\rho_1$ must be a function of $\tau$ if $I_2'(\psi) \neq 0$.
Instead of demanding this relation for $\rho_1$ (cf. Remark\,\ref{remark:generalization}),
we recourse to an assumption $I_2'(\psi) =0$
(the implication of this simple condition will be
discussed later).
Then, (\ref{Beltrami3'}) reduces to
\begin{equation}
(\tau^2+1) \psi'' + 3\tau \psi' - (\tau^2+1)g'\psi' = 0.
\label{Beltrami3''}
\end{equation}
We note that the Beltrami condition (\ref{Beltrami3''}) is freed
from $\rho_\parallel(\sigma)$. This fact merits in solving
(\ref{friction-balance-2}); see (\ref{friction-balance-3}).

\begin{figure}
\begin{center}
\includegraphics[scale=0.7]{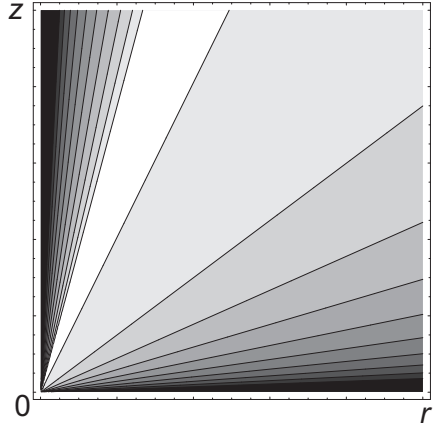}
\caption{ The distribution of $\rho_{\perp}$ of the similarity
solution (with $D=1$, $p=1$, $J=0.1$ and $q=1$). Contour levels
are given in the log scale of $\rho_{\perp}$.
}
\label{fig:density_1}
\end{center}
\end{figure}

For the specific form (\ref{similarity-1}) of $\psi$, we have to
determine an appropriate $g=\log \rho_\perp$ to satisfy
(\ref{Beltrami3''}), i.e.,
\begin{eqnarray}
g' &=&
\frac{\psi''}{\psi'} + \frac{3\tau}{\tau^2 + 1}
\nonumber
\\
&=&  \frac{J\,p\,(p-1)\tau^{p+q} +
D\,q\,(q+1)}{J\,p\,\tau^{p+q+1}-D\,q\,\tau}
+\frac{3\tau}{\tau^2+1} .
\label{similarity-3}
\end{eqnarray}
Solving (\ref{similarity-3}), we obtain
\[
g := \log\rho_\perp =
\log\frac{|Jp\tau^{p+q}-Dq|}{\tau^{q+1}}
+\frac{3}{2}\log(\tau^2+1) ,
\]
and, thus,
\begin{equation}
\rho_\perp = \frac{(\tau^2+1)^{3/2}|Jp\tau^{p+q}-Dq|}{\tau^{q+1}} .
\label{similarity-4}
\end{equation}
In Fig.\,\ref{fig:density_1}, we show the profile of $\rho_\perp(\tau)$.

\begin{remark}
\label{remark:generalization}
Here we considered the case of $I_2'(\psi) =0$, but a more
general solution can be obtained by demanding
$r^{-2}\rho_1$ to be a function of $\tau$.
For example, we may put $\rho_1=\rho_\perp(\tau)\rho_\parallel(\sigma)$
with $\rho_\parallel(\sigma)=\sigma^2=r^2+z^2$.  Then, $r^{-2}\rho_1=
(1+\tau^2)\rho_\perp(\tau)$.
\end{remark}

\subsection{Bernoulli relation in the disk region}
\label{subsec:disk_Bernoulli}

As mentioned above, this solution assumes $I_2'(\psi)
~(=\lambda)=0$, and hence, $I_2 = \rho_2 r V_\theta$ must
uniformly distribute. In the disk region (the vicinity of $z=0$),
we may approximate $V_\theta \approx \sqrt{MG/r}$ (Keplerian
velocity). Hence,  $\rho_2 \propto r^{-1/2}$.
In Fig.\,\ref{fig:density}, we show the profile of $\rho=\rho_1\rho_2$
for the case of $\rho_1\propto\rho_\perp$ (i.e., $\rho_\parallel=$constant).

For $\rho=\rho_1\rho_2=\rho_\parallel \rho_\perp r^{-1/2}$,
(\ref{friction-balance-2}) reads as
\begin{equation}
\nu = \frac{-\rmd\psi/\rmd z}{2 \rho_\parallel \rho_\perp} r^{-3/2}
\propto \frac{r^{-5/2}}{\rho_\parallel(r)}
\label{friction-balance-3}
\end{equation}
along each streamline in the disk region. For a given $\nu$, we
can solve (\ref{friction-balance-3}) for $\rho_\parallel$ to
determine the density profile.

In the disk region the Bernoulli relation (\ref{Bernoulli-1})
accounts as follows: by $\nabla\cdot\bm{P}=0$, we have $P_r =
\rho V_r \propto r^{-1}$.
If $\rho_\parallel(\sigma) =$ constant, for example,
$\rho\propto r^{-1/2}$ (evaluated along a streamline in the disk region).
Then, we have $V_r = V_{r0} r^{-1/2}$ with a (negative) constant $V_{r0}$.  Combining
the azimuthal velocity $V_\theta = V_{\theta 0} r^{-1/2}$ (which
must be slightly smaller than the Keplerian velocity
$\sqrt{MG/r}$), we obtain
\[
V^2 = (V_{r0}^2 + V_{\theta0}^2) r^{-1} = V_0^2 r^{-1}.
\]
By $\rho_2 \propto r^{-1/2}$, we obtain $ \partial_r (\log \rho_2)
= -(1/2) r^{-1}$. Hence, the Bernoulli relation
(\ref{Bernoulli-1}) demands
\[
\partial_r h = (V_{0}^2-MG) r^{-2},
\]
which yields $h = (MG -V_{0}^2) r^{-1}$. In this estimate, all
components of the energy density (gravitational potential $\phi$,
kinetic energy $V^2/2$ and enthalpy $h$) have a similar profile
($\propto r^{-1}$).

\subsection{Bernoulli relation in the jet region}
\label{subsec:jet_Bernoulli}
In the jet region (vicinity of $r=0$),
the streamlines (contours of $\tau=z/r$) are almost vertical, and
we may approximate $\sigma \approx z$.

Let us first estimate $\rho_2$ using
(\ref{friction-balance-2}), which is approximated, in the jet
region, by
\begin{eqnarray}
r \rho \nu = \{\log\rho_2,\psi\}
&\approx& (\partial_r\psi)\,(\partial_z\log\rho_2)
\nonumber \\
&=& J\,p\,\tau^{p+1}\frac{1}{z}\,
\partial_z(\log\rho_2), \label{jet-Bernoulli-1}
\end{eqnarray}
which shows that $\rho_2$ is an increasing function of $|z|$.
Using $\rho = \rho_\parallel(\sigma)\rho_\perp(\tau)\rho_2$, we
integrate (\ref{jet-Bernoulli-1}) along the streamline
($\tau=$constant, $\sigma \approx z$):
\begin{equation}
\frac{\rmd\rho_2}{\rho_2^2} = - \rmd\left(\frac{1}{\rho_2}\right)
=\frac{\nu}{Jp}\ \tau^{p+2}\,\rho_\perp(\tau) \rho_\parallel(z) z^2
\rmd z . \label{jet-Bernoulli-2}
\end{equation}
With this $\rho_2(z)$, we may estimate the toroidal (azimuthal)
component of the velocity: $V_\theta = I_2/(r\rho_2) =
(I_2\tau)/(z\rho_2)$, where $I_2$ and $\tau$ are constant (the
latter is constant along each streamline). We find that the
kinetic energy $V_\theta^2/2$ of the azimuthal velocity decreases
as a function of $|z|$ (both by the geometric expansion factor
$z^{-2}$ and the viscosity effect $\rho_2^{-2}$). The steep
gradient of the corresponding hydrodynamic pressure yields a
strong boost near the foot point ($z\approx 0$).

The poloidal component of the kinetic energy is estimated as follows:
We may approximate
\begin{eqnarray}
\frac{1}{2}(V_r^2 + V_z^2) = \frac{1}{2\rho^2 r^2}\,|\nabla\psi|^2
&\approx& \frac{1}{2\rho^2 r^2} \left(
Jp\frac{z^p}{r^{p+1}}\right)^2
\nonumber \\
&=&
\frac{(Jp)^2\tau^{2p+4}}{2\rho^2z^4} . \label{jet-Bernoulli-2'}
\end{eqnarray}
Here, the vertical distribution of the density $\rho =
\rho_\perp(\tau) \rho_\parallel(\sigma) \rho_2$ is primarily
dominated by $\rho_\parallel(\sigma)\approx\rho_\parallel(z)$.

At long distance from the origin, the jet has a natural similarity
property. For simplicity, let us ignore the effect of viscosity
($\nu=0$), and assume $\rho_2=1$. Then, $\rho_\parallel \propto
|z|^{-3/2}$ yields $(V_r^2 + V_z^2)/2 \propto z^{-1}$, which may
balance with the gravitational potential energy $\phi = -MG
|z|^{-1}$. Note, that the azimuthal component of the kinetic
energy disappears at large scale ($V_\theta^2 \propto z^{-2}$).
The Bernoulli condition (\ref{Bernoulli-1}) gives $h$ that also
has a similar distribution of $\propto |z|^{-1}$.

\begin{figure} 
\begin{center}
\includegraphics[scale=0.8]{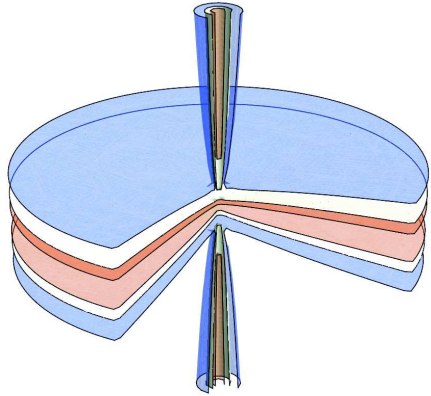}
\caption{
The contours (in log scale) of the density $\rho$
in the similarity solution (\ref{similarity-1})
of the Beltrami model (with parameters
$D=1$, $p=1$, $J=0.1$, and $q=1$).
}
\label{fig:density}
\end{center}
\end{figure}

Figure\,\ref{fig:density} shows the profile of $\rho =
\rho_\perp\rho_\parallel\rho_2$ with $\rho_\parallel \propto
|z|^{-3/2}$ (jet region) and $\rho_2 \propto r^{-1/2}$ (disk
region).
Dividing $\bm{P}$ by the density $\rho$,
we obtain the velocity field $\bm{V}$;  Fig.\,\ref{fig:streamline}
shows the streamlines of $\bm{V}$ corresponding to Figs.\,\ref{fig:flow}
and \ref{fig:density}.

\section{Summary and Concluding Remarks}

We have shown that the combination of a thin disk and
narrowly-collimated jet is the unique structure that is
amenable to the singularity of the Keplerian vorticity; the
Beltrami condition ---the alignment of flow
and \emph{generalized vorticity}--- characterizes the geometry.
Here the conventional vorticity is generalized as (\ref{g-vorticity}) to
subtract the viscosity force causing the accretion
and the centrifugal force of the Keplerian velocity.
Identifying the disk-jet structure as a generalized Beltrami vortex, we will be
able to understand the self-organization process in terms of the
``generalized helicity.''
As we have learned in the present practice,
the helicity of the generalized vorticity is the key parameter
that characterizes the self-organizing of a disk-jet system.

We end this paper with a short comment on the singularity of the Keplerian velocity.
The similarity solution has a singularity at the origin (where
$\phi=-MGr^{-1}$ diverges), which disconnects the
disk and jet parts of our solution.
To ``connect'' both subsystems, we need a \emph{singular
perturbation} that avoids the divergence of physical quantities
dictating the small-scale hierarchy on which the disk and jet regions
are connected smoothly;
Shiraishi {\it et al.}\,\cite{bib:Shiraishi} describes how the different topologies of magnetic
field-lines in the disk and jet regions can be connected in a
``boundary layer'' determined by the the Hall effect in
a weakly ionized plasma.

\section*{Acknowledgments}

The authors are grateful to Professor S. M. Mahajan,
Professor R. Matsumoto, Professor G. Bodo,
and Professor V. I. Berezhiani for
their suggestions and comments.
The work of ZY was supported by
Grant-in-Aid for Scientific Research (23224014) from MEXT, Japan,
and that of NLS was partially supported by the Rustaveli NSF Grant
project 1-4/16 (GNSF/ST09-305-4-140).

\appendix
\section{MHD model}
While this article describes a pure fluid-mechanical model of jet collimation,
many authors invoke a magnetic field, thrusting the center of the disk,
to ``guide'' (and twist, as often observed) the flow of charged gas (plasma).
Here the fluid vorticity plays the same role of a magnetic field.
Indeed, the vorticity of the \emph{canonical momentum}
combines the fluid vorticity and the magnetic field:
$\nabla\times(m\bm{V}+q\bm{A})$ ($m$ is the mass, $q$ is the charge,
and $\bm{A}$ is the vector potential of electromagnetic field).
Using this ``canonical vorticity,''
we may readily extend the present model to include the effect of
magnetic field
(in a Keplerian system, however, the singularity of the fluid vorticity may be the
principal part of the canonical vorticity).

The coupling of flow and magnetic field is described by the
magnetohydrodynamic (MHD) equations:
\begin{eqnarray}
& &
\partial_t \bm{P} + \nabla\cdot(\bm{V}\bm{P})
= (\nabla\times\bm{B})\times\bm{B} -\rho\nabla\phi - \nabla p
-\nabla\cdot\bm{\Pi},
\label{MHD:momentum-i-1}
\\
& &\partial_t \bm{B}= \nabla\times\left(\bm{V}\times\bm{B}\right),
\label{MHD:momentum-e-1}
\end{eqnarray}
where $\bm{B}$ is the magnetic field
(the magnetic energy density $|\bm{B}|^2/8\pi$ is normalized by
the kinetic energy density ${\cal{E}}_0:=\rho_0 V_0^2/2$).

A stationary solution of (\ref{MHD:momentum-e-1}) is given by
\begin{equation}
\bm{B} = \mu \bm{P}, \label{Beltrami-0}
\end{equation}
where $\mu$ is a certain scaler function (representing the
reciprocal Alfv\'en Much number). While (\ref{Beltrami-0}) is not
a general solution, other solutions are possible only if
the electron pressure $p_e$ or the electrostatic potential $\varphi$ is huge
(of the order of the kinetic energy density ${\cal E}_0$);
a perpendicular component of $\bm{B}$ with
respect to $\bm{P}$ causes a Lorentz force on electrons,
which must be balanced by a potential force $\nabla (p_e - \varphi)$.
If these energy densities are small, $\bm{B}$ of order unity is only
possible in the parallel direction of $\bm{P}$.
Operating divergence on both sides of (\ref{Beltrami-0}),
we find $\nabla\cdot(\mu \bm{P}) = \bm{P}\cdot\nabla\mu=0$,
implying $\mu=\mu(\psi)$.

Adding the magnetic field, the static force balance equation
(\ref{momentum-i-1-scaled}) is generalized as
\begin{equation}
\nabla\cdot(\bm{V}\bm{P}) - [\nabla\times(\mu\bm{P})]\times(\mu\bm{P})
= -\rho\nabla\phi - \nabla p - \nabla\cdot\bm{\Pi}.
\label{MHD:momentum-i-1-scaled}
\end{equation}
The \emph{generalized vorticity} is now combined with the magnetic field as
\begin{equation}
\widetilde{\bm{\Omega}_2}
:= \nabla\times\bm{P}_2  - \mu\rho_2\nabla\times\bm{B} .
\label{MHD:g-vorticity}
\end{equation}
This generalized vorticity is compared with that of the Hall-MHD
theory, i.e. the canonical
vorticity\,\cite{bib:MahajanYoshida1998}. Here the combination of the
mechanical and electromagnetic components are scaled by the physical
parameters $\mu$ (measuring the magnitude of the magnetic field)
and $\rho_2$ (reducing the mechanical component by the viscosity force).
The Beltrami equations (\ref{Beltrami-I-unmag}) and (\ref{Beltrami3'}) are,
respectively, generalized as
\begin{eqnarray}
& & \lambda \rho_1^{-1} \nabla\psi= \nabla \left(\rho_1^{-1}I \right)
-\mu\rho_2 \nabla \left( \mu I \right) ,
\label{MHD:Beltrami-I}
\\
& & \left(\rho_1^{-1}-\mu^2\rho_2 \right){\cal L}\psi
+ \nabla\psi\cdot\left(\nabla\rho_1^{-1}
-\mu\rho_2\nabla\mu \right)
= -\lambda \rho_1^{-1} I.
\label{MHD:Grad-form-3}
\end{eqnarray}
Combining (\ref{MHD:Beltrami-I}) and (\ref{MHD:Grad-form-3}), we obtain
\begin{equation}
{\cal L}\psi-\nabla\psi\cdot\nabla I
= \lambda\frac{|\psi|^2+I_2 I}{I(1-\mu^2\rho)} ,
\label{MHD:Grad-form-4}
\end{equation}
which reduces into (\ref{Beltrami3'}) when $\mu=0$ (i.e. unmagnetized).


\end{document}